\documentclass[aps]{revtex4}
\usepackage[dvips]{graphicx}
\usepackage{amsmath}

\begin{document}
\title{Monopole-like Configuration from Quantized SU(3)
Gauge Fields}
\author{V. Dzhunushaliev}
\email{dzhun@hotmail.kg}
\affiliation{Phys. Dept., Kyrgyz-Russian Slavic University,
Bishkek, 720000, Kyrgyz Republic}
\author{D. Singleton}
\email{dougs@csufresno.edu}
\affiliation{Physics Dept., CSU Fresno, 2345 East San Ramon Ave.
M/S 37 Fresno, CA 93740-8031, USA}

\begin{abstract}
Monopole field configurations have been extensively studied in
both Abelian and non-Abelian gauge theories. The question of
the quantum corrections to these systems is a difficult one,
since the classical monopoles have non-perturbatively large
couplings, which makes the standard, perturbative methods for
calculating quantum corrections suspect. Here we apply a
modified version of Heisenberg's quantization technique for strongly
interacting, nonlinear fields to a classical solution of the
SU(3) Yang-Mills field equations. This classical solution is not
monopole-like and has an energy density which diverges as $r
\rightarrow \infty$. However, the quantized version of this solution
has a monopole-like far field, and a non-divergent
energy density as $r \rightarrow
\infty$. This may point to the conclusion that
monopoles may arise not from quantizing classical
monopole configurations, but from quantizing field
configurations which at the classical level do not appear
monopole-like.
\end{abstract}


\maketitle

\section{Introduction}
Monopoles have been studied within the context of both
Abelian \cite{dirac} and non-Abelian \cite{thooft} gauge
theories. Usually, the approach is to start with a
classical monopole configuration ({\it i.e.} having a
magnetic field which becomes Coulombic as $r \rightarrow
\infty$) and then consider the quantum corrections
to the system. One difficulty with this approach is
that the coupling strength of monopole configurations
is large due to the Dirac quantization condition \cite{dirac}
which requires that there be an inverse relationship between
electric and magnetic couplings. A perturbatively small
electric coupling requires a non-perturbatively large
magnetic coupling. The perturbative quantization techniques
do not work well with monopoles for the same reason that they
have trouble with QCD in the low energy regime : the couplings
are non-perturbatively large.

Here we will use a modification of a non-perturbative
quantization method originally used by Heisenberg to
quantize nonlinear, strongly coupled spinor fields
\cite{hs1} \cite{hs3}. However, rather than applying
this quantization to an SU(3) field configuration
which is monopole-like already at the classical level,
we apply it to a non-monopole configuration, which
has fields and an energy density which diverge
at spacial infinity ($r \rightarrow \infty$). After
applying Heisenberg's quantization technique to this
solution we find that the fields and
energy density become physically well behaved, and
the asymptotic magnetic field becomes monopole-like. This
result may indicate that monopoles are inherently
quantum objects. Rather than quantizing
a classical field theory configuration which appears
monopole-like, real monopoles might arise as a
consequence of quantizing non-monopole classical solutions.

\section{Classical SU(3) solution}

In this section we will briefly review the classical,
spherically symmetric, SU(3) Yang-Mills theory solution \cite{dzh1}
to which we will apply the modified Heisenberg quantization
method. We begin with the following ansatz for pure SU(3) Yang-Mills
theory
\begin{eqnarray}
A _0 & = & \frac{1}{2}\lambda ^a
      \left( \lambda ^a _{ij} + \lambda ^a_{ji} \right )
      \frac{x^ix^j}{r^2} w(r),
\label{s1-1}\\
A^a_i & = & \left( \lambda ^a_{ij} - \lambda ^a_{ji} \right )
        \frac {{\bf i} x^j}{r^2} +
        \lambda ^a_{jk} \left (\epsilon _{ilj} x^k +
    \epsilon _{ilk} x^j\right ) \frac{x^l}{r^3} v(r)
\label{s1-2},
\end{eqnarray}
here $\lambda ^a$ are the Gell - Mann matrices; $a=1,2,\ldots ,8$
is a color index; the Latin indices $i,j,k,l=1,2,3$ are space indices;
${\bf i}^2=-1$; $r, \theta, \varphi$ are the usual spherically coordinates.
This is a simplified version of an ansatz considered by several
groups \cite{dzh1} -- \cite{galt}.
\par
Under this ansatz the Yang - Mills equations
($D^{\mu} {F^a}_{\mu \nu}= 0$) become
\begin{eqnarray}
r^2 v''& = & v^3 - v - vw^2,
\label{s1-3}\\
r^2 w''& = & 6w v^2.
\label{s1-4}
\end{eqnarray}
where the primes indicate differentiation with respect to $r$.
In the asymptotic limit $r \rightarrow \infty$ the solutions
to Eqs. \eqref{s1-3}, \eqref{s1-4} approach the form
\begin{eqnarray}
v & \approx & A \sin \left (x^{\alpha } + \phi _0\right ),
\label{s1-5}\\
w & \approx & \pm\left [ \alpha  x^{ \alpha } +
\frac{\alpha -1 }{4}\frac{\cos {\left (2x^{\alpha} + 2\phi _0 \right )}}
{x^{\alpha}}\right ],
\label{s1-6}\\
3A^2 & = & \alpha(\alpha - 1)
\label{s1-7}
\end{eqnarray}
where $x=r/r_0$ is a dimensionless radius and $r_0, \phi _0$, and $A$ are
constants. When Eqs. \eqref{s1-3}, \eqref{s1-4} were studied
numerically \cite{dzh1} it was found that the approximate form of the
solution in Eqs. \eqref{s1-5} - \eqref{s1-7} was a good representation
even for $x$ fairly close to the origin. It was also found that for a
broad range of boundary conditions for the functions $v, w$ that
$\alpha$ tend to fall in the range $1.2 \le \alpha \le 1.8$. Thus
for large $x$ the ansatz function $w$ is a smoothly, increasing
function, which yields an increasing time-like gauge potential
from Eq. \eqref{s1-1}. This can be compared with the confining
potentials used in some phenomenological studies of
quarkonia bound states \cite{eich}. It is also possible
to find analytic solutions with such increasing
gauge potentials if one couples the Yang-Mills fields
to a scalar field \cite{yos}. However, this
increasing of $w$ also leads to the field energy and action
of this solution diverging as $x \rightarrow \infty$.
\par
The ``magnetic'' ($H_i ^a = F^a _{jk}$) and ``electric''
($E_i ^a = F^a _{0i}$) fields associated with this solution
can be found from the non-Abelian gauge potentials, $A_{\mu} ^a$,
and have the following behavior
\begin{alignat}{3}
H^a _r & \propto & \frac{v^2-1}{r^2} ,
& \qquad H^a_{\varphi}  \propto  v' ,
& \qquad H^a_{\theta}  \propto  v' ,
\label{s1-8}\\
E^a_r & \propto & \frac{rw' - w}{r^2},
& \qquad E^a_{\varphi} \propto  \frac{vw}{r},
& \qquad E^a_{\theta}  \propto  \frac{vw}{r},
\label{s1-9}
\end{alignat}
here for $E^a _r , H^a _{\theta}$, and $H^a_{\varphi}$ the color index
$a=1,3,4,6,8$ and for $H^a _r, E^a _{\theta}$ and $E^a _{\varphi}$
$a=2,5,7$.

\section{Heisenberg's Non-Perturbative Quantization}

Although the confining behavior of the above classical solution
is of interest due to its similarity with certain phenomenological
potentials, its infinite field energy makes
its physical importance/meaning uncertain.
One possible resolution to this classical solution's
divergent field energy is if quantum effects
removed its bad long distance behavior. The difficulty is that
strongly interacting, nonlinear theories are notoriously hard to
quantize. In order to take into account the quantum effects
on this solution we will employ a variation of the method
used by Heisenberg \cite{hs1} in attempts to
quantize the nonlinear Dirac equation. We will outline
the key points of the method by using the nonlinear
Dirac equation as an illustrative example.
The nonlinear spinor field equation considered by
Heisenberg had the following form :
\begin{equation}
\label{hq-1}
\gamma ^{\mu} \partial _{\mu} {\hat \psi (x)} - l^2
\Im [{\hat \psi} ({\hat {\bar \psi}}
\psi) ] = 0
\end{equation}
where $\gamma ^{\mu}$ are Dirac matrices; ${\hat \psi (x)},
{\hat {\bar \psi}}$ are the spinor field and its adjoint
respectively; $\Im [{\hat \psi} ({\hat {\bar \psi}} {\hat \psi}) ]$
is the general nonlinear spinor self interaction term which
involved three spinor fields and various combinations of $\gamma^{\mu}$'s
and/or $\gamma ^5$'s. The constant $l$ has units of length, and sets the
scale for the strength of the interaction. Next one defines $\tau$
functions as
\begin{equation}
\label{hq-2}
\tau (x_1 x_2 ... | y_1 y_2 ...) = \langle 0 | T[{\hat \psi} (x_1)
{\hat \psi} (x_2) ... {\hat \psi} ^{\ast}
(y_1) {\hat \psi} ^{\ast} (y_2) ...] | \Phi \rangle
\end{equation}
where $T$ is the time ordering operator; $| \Phi \rangle$ is a state for
the system described by Eq. \eqref{hq-1}. Applying
Eq. \eqref{hq-1} to \eqref{hq-2} we obtain the following infinite
system of equations for various $\tau $'s
\begin{eqnarray}
\label{hq-3}
l^{-2} \gamma ^{\mu} _{(r)} \frac{\partial}{\partial x^{\mu} _{(r)}}
&&\tau (x_1 ...x_n |y_1 ... y_n ) = \Im [ \tau (x_1 ... x_n x_r |
y_1 ... y_n y_r)] + \nonumber \\
&&\delta (x_r -y_1) \tau ( x_1 ... x_{r-1} x_{r+1} ... x_n |
y_2 ... y_{r-1} y_{r+1} ... y_n ) + \nonumber \\
&&\delta (x_r - y_2) \tau (x_1 ... x_{r-1} x_{r+1} ... x_n |
y_1 y_2 ... y_{r-1} y_{r+1} ... y_n ) + ...
\end{eqnarray}
Eq. \eqref{hq-3} represents one of an infinite set of coupled equations
which relate various order (given by the index $n$) of the $\tau$
functions to one another. To make some head way toward solving
this infinite set of equations Heisenberg employed the Tamm-Dankoff
method whereby he only considered $\tau$ functions up to a certain
order. This effectively turned the infinite set of coupled equations
into a finite set of coupled equations.
\par
For the SU(3) Yang-Mills theory this idea leads to the following
Yang-Mills equations for the quantized SU(3) gauge field
\begin{equation}
  D_\mu \hat F^{a\mu\nu} = 0
  \label{hq3a}
\end{equation}
here $\hat F^{a\mu\nu}$ is the field operator of the SU(3)
gauge field.
\par
One can show that Heisenberg's method
is equivalent to the Dyson-Schwinger system of equations for
small coupling constants. One can also make a comparison
between the Heisenberg method and the standard Feynman
diagram technique. With the Feynman diagram method quantum
corrections to physical
quantities are given in terms of an infinite number of higher
order, loop diagrams. In practice one takes only a finite
number of diagrams into account when calculating the quantum
correction to some physical quantity.
This standard diagrammatic method requires a small
expansion parameter (the coupling constant), and thus
does not work for strongly coupled theories. The Heisenberg
method was intended for strongly coupled, nonlinear theories,
and we will apply a variation of this method to the classical
solution discussed in the last section.

We will consider a variation of Heisenberg's quantization method
for the present non-Abelian equations by making the following
assumptions \cite{dzh2}:
\begin{enumerate}
\item
The physical degrees of freedom relevant for studying
the above classical solution
are given entirely by the two ansatz
functions $v, w$ appearing in Eqs. \eqref{s1-3}, \eqref{s1-4}.
No other degrees of freedom will arise through
the quantization process.
\item
From Eqs. \eqref{s1-5}, \eqref{s1-6} we see that one function
$w(r)$ is a smoothly varying function for large
$r$, while another function, $v(r)$, is strongly oscillating.
Thus we take $w(r)$ to be an almost
classical degree of freedom while
$v(r)$ is treated as a
fully quantum mechanical degree of freedom.
Naively one might expect that
only the behavior of second function
would change while first function stayed
the same. However since both functions are interrelated
through the nonlinear nature of the field equations
we find that both functions are modified.
\end{enumerate}
To begin we replace the ansatz functions by operators ${\hat w} (x) ,
{\hat v} (x)$.
\begin{eqnarray}
\label{s3-1}
r^2 {\hat v}'' &=& {\hat v}^3 - {\hat v} - {\hat v} {\hat w} ^2 ,
\label{s3-2} \\
r^2 {\hat w}'' &=& 6{\hat w} {\hat v}^2
\label{s3-3}
\end{eqnarray}
These equations can be seen as an approximation of
the quantized SU(3) Yang-Mills field equations \eqref{hq3a}.
Taking into account assumption (2) we let
${\hat w} \rightarrow w$ become just
a classical function again, and replace
${\hat v}^2$ in Eq. \eqref{s3-3}
by its expectation value to arrive at
\begin{eqnarray}
r^2 {\hat v}'' &=& {\hat v}^3 - {\hat v} - {\hat v} w ^2 ,
\label{s3-4} \\
r^2 w'' &=& 6 w \langle v^2 \rangle
\label{s3-5}
\end{eqnarray}
where the expectation value $\langle {\hat v}^2 \rangle$ is taken with
respect to some quantum state $|q\rangle$:
$\langle v^2 \rangle = \langle q| v^2 | q\rangle $.
We can average Eq. \eqref{s3-4} to get
\begin{eqnarray}
  r^2 \langle v \rangle '' & = &
  \langle  v^3 \rangle - \langle v \rangle -
  \langle v \rangle w^2 ,
  \label{s3-6}\\
  r^2 w'' & = & 6 w \langle v^2 \rangle
  \label{s3-7}
\end{eqnarray}
Eqs. \eqref{s3-6} \eqref{s3-7} are almost a closed system for
determining $\langle v \rangle$ except for the $\langle v^2 \rangle$
and $\langle v^3 \rangle$ terms. One can obtain differential equations
for these expectation values by applying $r^2 \partial
/ \partial r$ to ${\hat v}^2$ or ${\hat v}^3$ and using
Eqs. \eqref{s3-4} - \eqref{s3-5}. However the differential
equations for $\langle v^2 \rangle$ or $\langle v^3 \rangle$
would involve yet higher powers of ${\hat v}$ thus generating
an infinite number of coupled differential equations for the various
$\langle v^n \rangle$. In the next section we will use a path integral
{\em inspired} method \cite{dzh3} to cut this progression off at some finite
number of differential equations.

\section{Path integration over classical solutions}

Within the path integral method the expectation value of
some field $\Phi$ is given by
\begin{equation}
\label{av-2}
  \langle \Phi \rangle = \int  \Phi e^{i S
  \left [  \Phi \right ]} D \Phi
\end{equation}
The classical solutions, $\Phi _{cl}$, give the dominate
contribution to the path integral. For a single classical
solution one can approximate the path integral as
\begin{equation}
  \int e^{iS[\Phi]} D\Phi \approx A e^{iS[\Phi_{cl}]}
  \label{av-2a}
\end{equation}
where $A$ is a normalization constant. Consequently the
expectation of the field can be approximated by
\begin{equation}
  \int \Phi e^{iS[\Phi]} D\Phi \approx \Phi_{cl} .
  \label{av-3a}
\end{equation}
We are interested in the case where $\Phi$ is the gauge
potential $A^a_\mu$, in which case our approximation becomes
\begin{equation}
  \langle A^a_\mu \rangle \approx
  \int \left (\tilde A^a_\mu \right )_{\phi _0}
  e^{i S  \left [ \left (\tilde A^a_\mu
  \right )_{\phi _0} \right ]}
  D \left (\tilde A^a_\mu \right )_{\phi _0}
  \label{av-1}
\end{equation}
$(\tilde A^a_\mu)_{\phi _0}$  are the classical
solutions of the Yang - Mills equations labeled by a parameter
$\phi _0$. In the present case the classical solution with
the asymptotic form \eqref{s1-5} \eqref{s1-6} has an infinite
energy and action. When one considers the Euclidean version
of the path integral above the exponential factor in
\eqref{av-1} becomes $\exp[-S [ (\tilde A^a_\mu
)_{\phi_0}]$ which for an infinite action
would naively imply that this classical configuration
would not contribute to the path integral at all. However,
there examples where infinite action classical solutions
have been hypothesized to play a significant role in the
path integral. The most well known example of this is
the meron solution \cite{callan}, which has an infinite
action. Analytically the singularities of the meron
solutions can be dealt with by replacing the regions that
contain the singularities by instanton solutions. Since
instanton solutions have finite action this patched
together solution of meron plus instanton has finite action.
However, the Yang-Mills field equations are not
satisfied at the boundary where the meron and instanton
solutions are sewn together. In addition recent
lattice studies \cite{sn} \cite{negele}
have indicated that merons (or the patched meron/instanton)
do play a role in the path integral. Here we will treat
this divergence in the action in an approximate way through
a redefinition of the path integral integration measure.
Since the divergence in the action comes from first term
of the $w(r)$ ansatz function of \eqref{s1-6}, which does
not contain $\phi _0$ we will take the action for different
$\phi_0$ 's to be approximately the same : 
$S[(\tilde A^a_\mu )_{\phi_0} ] \approx S_0 \rightarrow \infty$.
Then changing the functional integration measure in
the following way -- $D(\tilde A^a_\mu )_{\phi_0} \rightarrow
e^{-iS_0}D(\tilde A^a_\mu )_{\phi_0}$ -- allows us to approximate
the expectation value of $A^a _{\mu}$ as
\begin{equation}
  \langle A^a_\mu \rangle \approx
  \sum _{\substack{\text{over all} \\
  \text{classical solutions}}}
  \left (
  \tilde {A^a_\mu}
  \right )
  _{\phi _0} p_{\phi _0}
  \label{av-1a}
\end{equation}
where $p_{\phi _0}$ is the probability for a given
classical solution. We will consider the classical
solutions whose asymptotic behavior is given by \eqref{s1-5}-\eqref{s1-6},
and we will take the different solutions (as distinguished
by different $\phi _0$'s) to have equal probability
$p_{\phi _0} \approx const$. Therefore
\begin{eqnarray}
  \langle v \rangle & \approx &
  \frac{1}{2\pi} \int\limits^{2\pi}_0 v_{cl} d\phi _0 =
  \frac{A}{2\pi}
  \int\limits^{2\pi}_0 \sin(x^\alpha + \phi _0) d\phi _0  = 0 ,
  \label{av-3}\\
  \langle v^2 \rangle & \approx &
  \frac{1}{2\pi} \int\limits^{2\pi}_0 v^2_{cl} d\phi _0 =
  \frac{A^2}{2\pi}  \int\limits^{2\pi}_0
  \sin^2 (x^\alpha + \phi _0) d\phi _0 =
  \frac{A^2}{2} ,
  \label{av-4}\\
  \langle v^3 \rangle & \approx &
  \frac{1}{2\pi} \int\limits^{2\pi}_0 v^3_{cl} d\phi _0 =
  \frac{A^3}{2\pi}  \int\limits^{2\pi}_0
  \sin^3 (x^\alpha + \phi _0) d\phi _0 = 0 ,
  \label{av-5}\\
  \langle v^4 \rangle & \approx &
  \frac{1}{2\pi} \int\limits^{2\pi}_0 v^4_{cl} d\phi _0 =
  \frac{A^4}{2\pi}  \int\limits^{2\pi}_0
  \sin^4 (x^\alpha + \phi _0) d\phi _0 =
  \frac{3}{8} A^4
  \label{av-6}
\end{eqnarray}
here $v_{cl}$ is the function from the Eq.\eqref{s1-5}.
The path integral {\em inspired} Eqns. \eqref{av-3} -
\eqref{av-6} are the heart of the cutoff procedure
that we wish to apply to Eqns. \eqref{s3-6}, \eqref{s3-7}.
On substituting Eqs. \eqref{av-3}, \eqref{av-4}
into Eqs. \eqref{s3-6}, \eqref{s3-7} we find that
Eq.\eqref{s3-6} is satisfied identically
and Eq.\eqref{s3-7} takes the form
\begin{equation}
  r^2 w'' = 3A^2w = \alpha (\alpha - 1) w ,
  \qquad \alpha >1
  \label{av-7}
\end{equation}
which has the solutions
\begin{eqnarray}
  w & = & w_0 r^\alpha ,
  \label{av-8}\\
  w & = & \frac{w_0}{r^{\alpha - 1}}
  \label{av-9}
\end{eqnarray}
where $w_0$ is some constant. The first
solution is simply the classically averaged singular solution
\eqref{s1-6} which still has the bad asymptotic divergence of the
fields and energy density. The second solution,
\eqref{av-9}, is more physically relevant since
it leads to asymptotic fields which are well behaved.
\par
The solution of Eq. \eqref{av-9} implies the following important
result: \textbf{\textit{the quantum fluctuations
of the strongly oscillating, nonlinear fields leads to an improvement
of the bad asymptotic behavior of these nonlinear fields.}}
This means that after quantization the monotonically growing
and strongly oscillating components of the gauge potential
become functions with good asymptotic behavior.
\par
As $r \rightarrow \infty$ we find the following SU(3) color fields
\begin{eqnarray}
  \langle H^a_r \rangle & \propto &
  \frac{\langle v^2 \rangle - 1}{r^2}
  \approx
  \frac{Q}{r^2}
  \qquad \text{with} \qquad
  Q = \frac{1}{6}\alpha (\alpha -1 ) - 1 ,
  \label{av-10}\\
  \langle H^a_{\varphi ,\theta} \rangle & \propto &
  \langle v' \rangle \approx 0 ,
  \label{av-11}\\
  \langle E^a_r \rangle & \propto &
  \langle \frac{rw' - w}{r^2} \rangle
  \approx
  -\frac{\alpha w_0}{r^{\alpha + 1}} ,
  \label{av-12}\\
  \langle E^a_{\varphi , \theta} \rangle & \propto &
  \frac{\langle v \rangle w}{r} = 0 .
  \label{av-13}
\end{eqnarray}
We can see that as $r \rightarrow \infty$
$|\langle E^a_r \rangle | \ll |\langle H^a_r \rangle |$.
In particular at infinity we find only a monopole
``magnetic'' field $H^a_r \approx Q/r^2$ with a ``magnetic''
charge $Q$. This result can be summarized as:
\textbf{\textit{the approximate quantization of the SU(3) gauge field
(by averaging over the classical singular solutions) gives a
monopole-like configuration from an initial classical configuration
which was not monopole-like.}} We will call this a
\textbf{\textit{``quantum monopole''}} to distinguish it
from field configurations which are monopole-like already in
the classical theory.

\section{Energy density}

The divergence of the fields of the classical solution
given by Eqs. \eqref{s1-5} - \eqref{s1-7} leads to a diverging
energy density for the solution, and thus an infinite
total field energy. The energy density $\varepsilon$ 
of the quantized solution is
\begin{equation}\label{en-0}
  \varepsilon \propto (E^a_\mu)^2 + (H^a_\mu)^2
   \propto \left( \frac{rg' - g}{r^2}\right)^2 +
   \frac{2\langle f^2\rangle g^2}{r^4} +
   \frac{2\langle {f'}^2\rangle}{r^2} +
   \frac{\langle (f^2 - 1)^2 \rangle}{r^4}
\end{equation}
The first two terms on the right hand side of Eq.
\eqref{en-0}, which involve the ``classical''
ansatz function $g(r)$, go to zero faster than
$1/r^4$ as $r \rightarrow \infty$ due to the form
of $g(r)$ in Eq. \eqref{av-9}. Thus the leading behavior
of $\varepsilon$ is given by the last two terms in
Eq. \eqref{en-0} which have only the ``quantum''
ansatz function $f(r)$. To calculate $\langle {f'}^2 \rangle$
let us consider
\begin{equation}
  r^2 \frac{d}{dr}
  \left\langle 
  f'(r') f'(r)
  \right\rangle = \left\langle f'(r') f^3(r) \right\rangle - 
  \left\langle f'(r') f(r) \right\rangle 
  \left( 1 + g^2(r) \right)
\label{en-1}
\end{equation}
In the limit $r' \rightarrow r$ we have
\begin{equation}
  r^2 \left\langle f'(r) f''(r) \right\rangle = 
  \frac{r^2}{2} \left\langle {f'}^2(r) \right\rangle ' = 
  \frac{1}{4} \left\langle f^4(r) \right\rangle ' - 
  \frac{1}{2} \left\langle f^2(r) \right\rangle ' 
  \left( 1 + g^2(r) \right).
\label{en-2}
\end{equation}
From Eqs. \eqref{av-4} and \eqref{av-6} we see that
$\langle {f'}^2 \rangle ' = 0$ which implies
$\langle {f'}^2 \rangle  = const.$
Thus the third term in \eqref{en-0} gives the leading
asymptotic behavior as $r \rightarrow \infty$ to be
\begin{equation}
  \varepsilon \approx \frac{const}{r^2}
  \label{en-3}
\end{equation}
and the total energy of this ``quantum monopole'' (excitation) 
is infinite. This fact indicates that our approximation \eqref{av-2} 
is good only for $\langle f^n(r) \rangle$ calculations but not for 
the derivative $\langle {f'}^2(r) \rangle$. .

\section{Conclusions}

\begin{figure}[htb]
\begin{center}
\fbox{
\includegraphics[height=5cm,width=8cm]{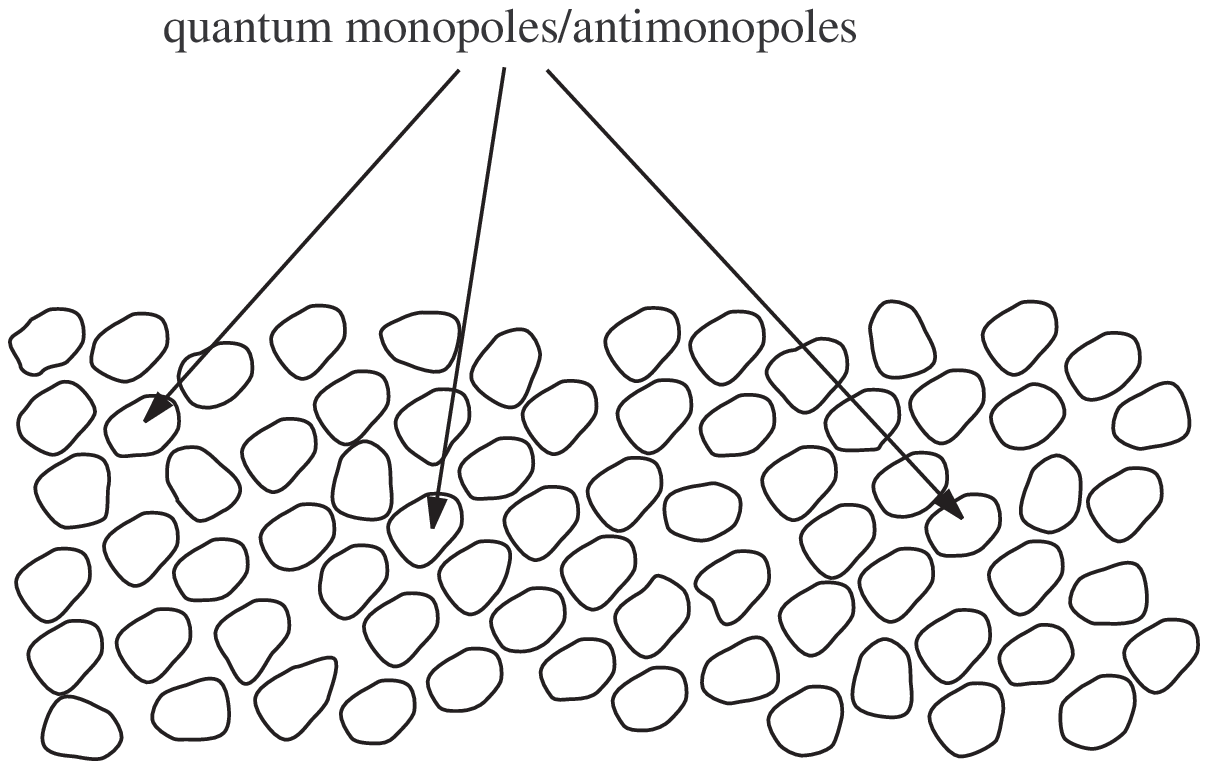}}
\caption{QCD vacuum $\approx$ stochastic gas of quantum
monopoles/antimonopoles.}
\label{fig1}
\end{center}
\end{figure}

Starting from an infinite energy, classical solution to the
SU(3) Yang-Mills field equations we found that the bad
asymptotic behavior of this solution was favorably modified by a
variation of the quantization method proposed by Heisenberg
to deal with strongly coupled, nonlinear field theories. In addition,
although the original classical solution was not monopole-like,
it was found that the quantized solution was monopole-like.
This may imply that if real monopoles exist they may be inherently
quantum mechanical objects {\it i.e.} that monopoles arise from
the quantization of non-monopole classical solutions. This is to be
contrasted with the standard idea that monopoles result from
quantizing solutions which are already monopole-like
at the classical level. One possible application of this
is to the dual-superconductor picture of the QCD vacuum. In this
picture one models the QCD vacuum as a stochastic gas of
appearing/disappearing monopoles and antimonopoles as in
Fig.\ref{fig1}. These monopole/antimonopole fluctuations can
form pairs (analogous to Cooper pairs in real superconductors)
which can Bose condense leading to a dual Meissner effect
\cite{hooft2} \cite{mand}
expelling color electric flux from the QCD vacuum,
expect in narrow flux tubes which connect and confine the
quarks. Lattice calculations confirm such a
model \cite{suzuki} : monopoles appear to play a
major role in the QCD lattice gauge path integral.
Based on the results of the present paper it may be that
the monopoles which are considered in the dual
superconductor QCD vacuum picture should be the
``quantum'' monopoles discussed here rather than ``classical''
monopoles ({\it i.e.} monopoles which are already monopole-like
at the classical level).

\section{Acknowledgment}

VD is grateful for Viktor Gurovich for the fruitful discussion.

\end{document}